\documentclass[aps,prb,letterpaper,10pt,twocolumn,notitlepage,superscriptaddress]{revtex4-1}
\usepackage{hyperref}
\usepackage{amssymb}
\usepackage{amsmath}
\usepackage{graphicx}
\usepackage{subfigure}
\usepackage{amsfonts}
\usepackage{xcolor}
\usepackage{epsfig}
\usepackage{color}
\usepackage{bbm}
\usepackage{bm}
\usepackage{tabularx}
\usepackage{multirow}
\usepackage{mathtools}
\usepackage{xfrac}
\usepackage{blkarray}
\usepackage{bbold}
\usepackage[mathscr]{euscript}
\usepackage{longtable}
\newcommand*{\ket}[1]{\vert{#1}\rangle} 
\newcommand*{\bra}[1]{\langle{#1}\vert} 
\newcommand*{\abs}[1]{\left|#1\right|}

\newcommand*{\vnnb}{V$_{\rm N}$N$_{\rm B}$}
\newcommand*{\vncb}{V$_{\rm N}$C$_{\rm B}$}
\newcommand*{\vn}{V$_{\rm N}$}
\newcommand*{\vb}{V$_{\rm B}$}
\newcommand*{\abinit}{\textit{ab initio} }

\newcommand*{\GSD}{\mathscr{B}^{0,\rm d}_{\pm\sfrac{1}{2}}}
\newcommand*{\FED}{\mathscr{A}^{1,\rm d}_{\pm\sfrac{1}{2}}}
\newcommand*{\SED}{\mathscr{B}^{2,\rm d}_{\pm\sfrac{1}{2}}}
\newcommand*{\TEQd}{\mathscr{A}^{3,\rm q}_{\pm\sfrac{1}{2}}}
\newcommand*{\TEQu}{\mathscr{A}^{3,\rm q}_{\pm\sfrac{3}{2}}}

\begin{document}

\title{Color centers in hexagonal boron nitride monolayers: A group theory and \textit{ab initio} analysis}

\author{Mehdi Abdi}
\email{mehabdi@gmail.com}
\affiliation{Institute of Theoretical Physics and IQST, Albert-Einstein-Allee 11, Ulm University, 89069 Ulm, Germany}
\affiliation{Department of Physics, Isfahan University of Technology, Isfahan 84156-83111, Iran}

\author{Jyh-Pin Chou}
\author{Adam Gali}
\affiliation{Wigner Research Centre for Physics, Hungarian Academy of Sciences, PO Box 49, 1525 Budapest, Hungary}
\affiliation{Department of Atomic Physics, Budapest University of Technology and Economics, Budafoki {\'u}t 8, 1111 Budapest, Hungary}
\author{Martin B. Plenio}
\affiliation{Institute of Theoretical Physics and IQST, Albert-Einstein-Allee 11, Ulm University, 89069 Ulm, Germany}


\begin{abstract}
We theoretically study physical properties of the most promising color center candidates for the recently observed single-photon emissions in hexagonal boron nitride (h-BN) monolayers.
Through our group theory analysis combined with density functional theory (DFT) calculations we provide several pieces of evidence that the electronic properties of the color centers match the characters of the experimentally observed emitters.
We calculate the symmetry-adapted multi-electron wavefunctions of the defects using group theory methods and analyze the spin-orbit and spin-spin interactions in detail.
We also identify the radiative and non-radiative transition channels for each color center.
An advanced \abinit DFT method is then used to compute energy levels of the color centers and their zero-phonon-line (ZPL) emissions.
The computed ZPLs, the profile of excitation and emission dipole polarizations, and the competing relaxation processes are discussed and matched with the observed emission lines.
By providing evidence for the relation between single-photon emitters and local defects in h-BN, this work provides the first steps towards harnessing quantum dynamics of these color centers.
\end{abstract}


\maketitle

%
%
\section{Introduction}
Thanks to the advances in fabrication and control, two-dimensional (2D) materials are under intense investigation nowadays for their potential applications in many areas including quantum technology~\cite{Britnell2012, Georgiou2013, He2015}. This ranges from quantum nanophotonics~\cite{Xia2014, Clark2016, Shiue2017} to quantum sensing~\cite{Lee2012, Abderrahmane2014, Li2014, Abdi2016} and quantum information processing~\cite{Cai2013,Tran2016a, Abdi2017}.
Recently, there has been an increasing interest in photon emitters in 2D materials~\cite{Srivastava2015, Chakraborty2015, Palacios2016}, in general, and layers of hexagonal boron nitride (h-BN), in particular~\cite{Aharonovich2016}. The recent reports on the observation of single photon emission from few layers of hexagonal boron nitride samples has sparked a considerable interest.
These experiments have provided a broad spectrum of data on the single-photon emitters from h-BN both in the visible~\cite{Tran2016a, Tran2016b, Chejanovsky2016, Jungwirth2016, Martinez2016, Schell2016, Shotan2016, Jungwirth2017, Li2017, Exarhos2017} and UV~\cite{Museur2008, Bourrellier2016, Vuong2016}. The origin of the emissions, however, is still under study. Nevertheless, based on the experimental observations and density functional theory (DFT) computations there are evidences leading to the local defects, the `color centers'~\cite{Wong2015, Tran2016b}.
Several color center candidates have already been suggested.
Despite considerable efforts by research made so far, detailed experimental investigations are still required to fully unveil the origin of these optical emissions~\cite{Jimenez1997, Museur2008, Jin2009}.

Color centers in 2D materials are envisioned to have applications in nanophotonics and, provided their electronic structure and magnetic properties are known well, they could be employed for many other purposes from quantum sensing to quantum information processing~\cite{Abdi2017}. Since they have the privilege of sitting in a 2D material, which means natural proximity to the surface their sensitivity to the surrounding environment is expected to be high. Their high quality zero-phonon-line (ZPL), on the other hand, makes them promising high quantum efficiency spin-photon interfaces.

A better understanding of the observed emissions from 2D materials and their origins is required for their control, which cannot be achieved without more theoretical investigations. The pioneering theoretical works are focused on the computational methods to study the stability and structural properties of h-BN flakes accommodating local defects~\cite{Orellana2001, Mosuang2002, Azevedo2007} as well as their electronic and magnetic properties~\cite{Si2007, Azevedo2009, Topsakal2009, Okada2009}.
Further investigations clarified that some defect families exhibit deep band-gap levels with partial occupation, which indicate allowed electric dipole transitions at optical frequencies and beyond~\cite{Attaccalite2011, Huang2012b}. Other first-principles considerations do not exclude dislocation lines, multi-vacancy sites, and topological defects like Stone-Wales as a resource of emission~\cite{Zobelli2006, Cretu2014, Yin2010, Wang2016}. However, the similarities between the h-BN optical emitters to the known color centers in other materials, e.g. nitrogen-vacancy (NV) and silicon-vacancy (SiV) centers in diamond~\cite{Aharonovich2016}, have tentatively brought the attention of the community to substitutional and vacancy defects. In particular, it is observed that the emitters can be created by ion bombardment in a controlled fashion~\cite{Choi2016}.
More recently, several DFT computational analyses have attributed the emissions to charge neutral native and substitutional defects with deep band-gap energy levels~\cite{Li2017, Tawfik2017, Cheng2017}. These reports lay the ground work for further research on the physics of such defects. But a group theory accompanied with DFT analysis is necessary for a deeper understanding of these color centers which can guide future experimental and theoretical works. A route that have proven to be beneficial in diamond color centers, e.g. NV and SiV centers~\cite{Doherty2011, Maze2011, Hepp2014}.

Nonetheless, to the best of our knowledge, there is no group theoretical investigation on these single photon emission candidates. To provide a better understanding on their electronic and magnetic properties, here we provide an analysis based on symmetry observations on a few of the most relevant candidates and take advantage of the observations made in the experiments and our \abinit calculations to explain such properties of the optical emitters via group theory analysis. By determining the symmetry-adapted total wavefunctions of the multi-electron states we present energy ordering of such states aided by our advanced DFT calculations. We perform thorough analyses on the effect of spin-orbit and spin-spin interactions as well as applied electric fields, as first inevitable step, at fixed coordinate of the atoms associated with the potential energy surface minimum at the given electronic configuration. The effect of electron-phonon coupling on the results will be  discussed briefly as well. These studies allows us to come to the conclusion that the defects we are investigating here correlate well with the observed emitter species in the experiments.
Remarkably, their electronic and magnetic properties allow for applications in quantum control and information processing~\cite{Cai2013}.

\begin{figure}[tb]
\includegraphics[width=0.7\columnwidth]{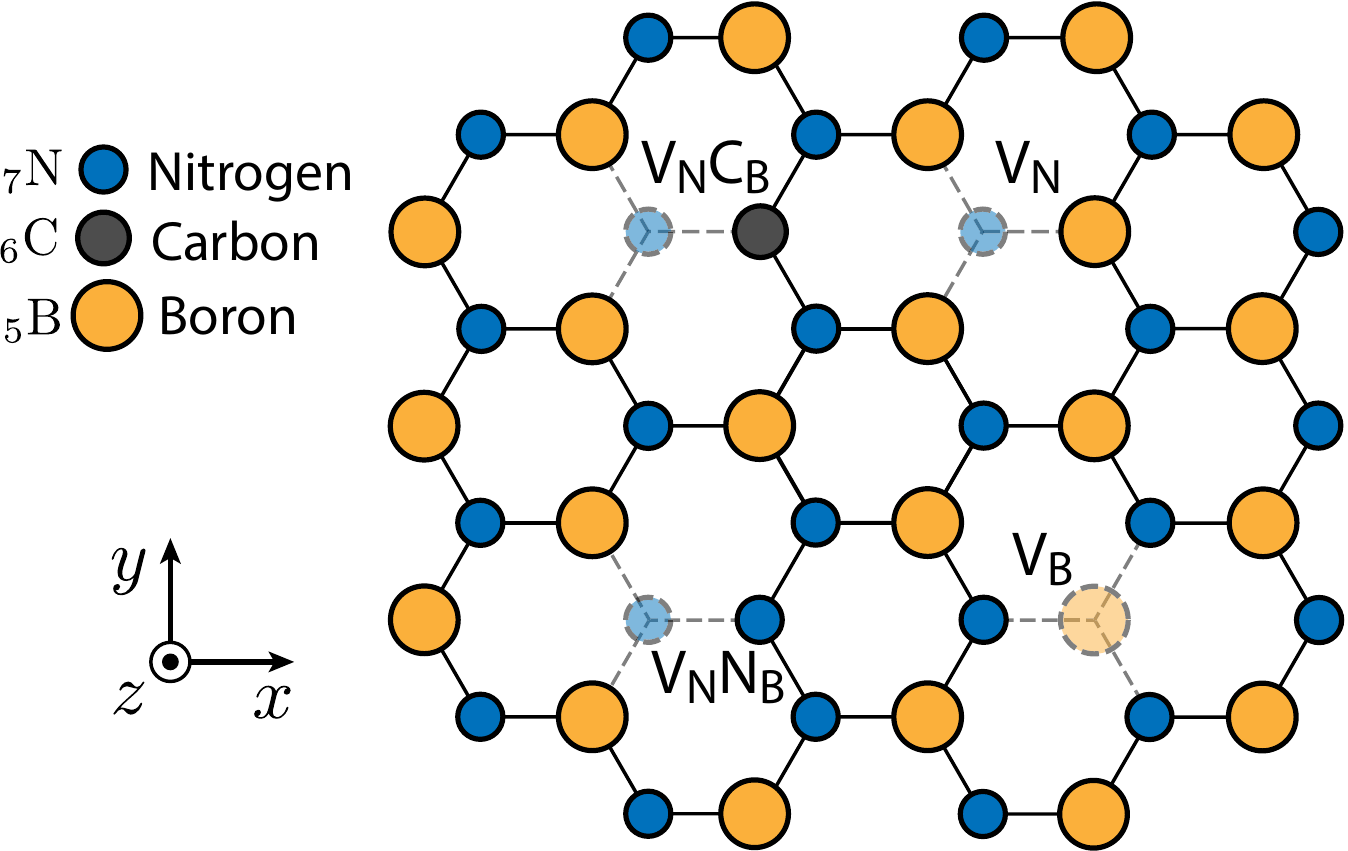}
\caption{The geometry of various possible defects in h-BN. Note that \vnnb\ and \vncb\ defects have $C_{2v}$ point group symmetry with their axis of symmetry laying in the plane ($x$-axis here), while the rest of defects have $D_{3h}$ point group symmetry with the symmetry axis pointing out of the plane ($z$-axis).}
\label{fig:scheme}
\end{figure}

Hexagonal boron nitride can accommodate a wide variety of local defects in its lattice structure.
This includes the most stable vacancy incorporated representatives: boron vacancy \vb, nitrogen vacancies \vn, a complex anti-site which is a nitrogen vacancy next to a nitrogen anti-site \vnnb, and substitutional anti-site defect like $\mathrm{V_NC_B}$ [see Fig.~\ref{fig:scheme}]. Our focus will be on the cases that can offer a platform for future quantum technological applications due to their nontrivial ground spin state. That is, those defects whose ground state is not a spin singlet. It is worth mentioning here that defects with electronic spin-singlet ground state can still be interesting provided their optical excited state has a nonzero spin state and in strong interaction with some neighboring spin systems, e.g. nuclear spins, though demanding more complicated control protocols.
The observations on optical single-photon emitters strongly supports the likelihood of defects with a vacancy. Furthermore, such defects are being created in a rather controllable way by ion irradiation~\cite{Grosso2017}. Therefore, we place our focus on defects with one vacancy. Observation of \vn\ and \vb\ defects is already reported in a TEM experimental work~\cite{Jin2009}, however, visual evidences on the existence of \vnnb\ and $\mathrm{V_NC_B}$ yet to come out.

In this paper, we provide a thorough group theoretical analysis combined with DFT calculations on the defects of our interest, namely the neutral \vnnb~and the negatively charged \vb.
The order of many-body levels are determined by our advanced DFT calculations.
Note that positively charged \vncb~as well as the singlet state situations resulting from the negatively and positively charged \vnnb~and neutral \vncb~are not excluded from being the source of single photon emission. Therefore, their electronic diagram and spin properties are discussed in the supporting information.

%
%
\section{Neutral \vnnb}
\begin{figure*}[tb]
\includegraphics[width=\textwidth]{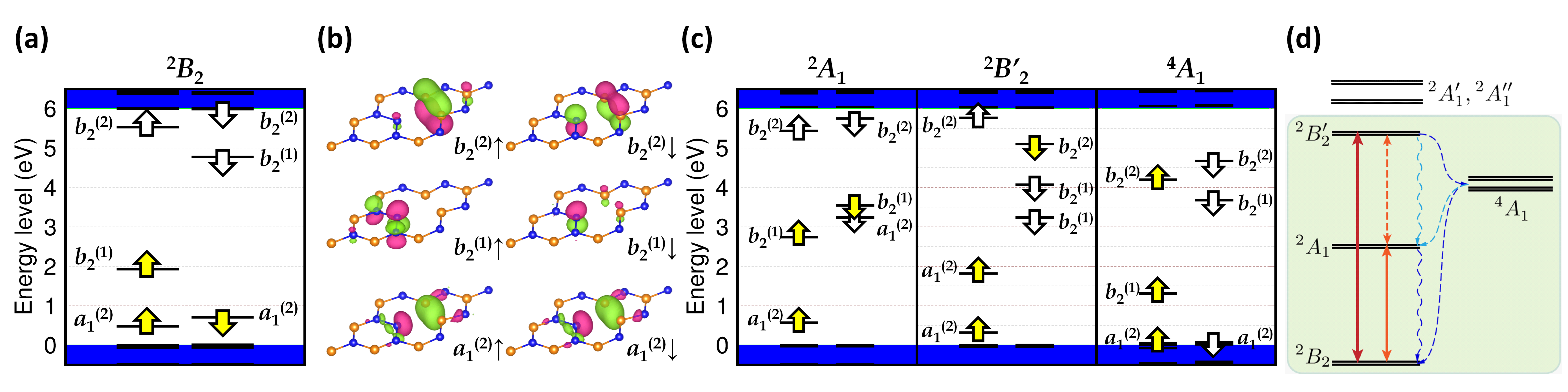}
\caption{%
(a) Single particle defect levels of the ground state neutral \vnnb~in the h-BN fundamental band gap. The orbitals labelled according to the group theoretical symmetry orbitals. The yellow and white arrows represent the occupied and unoccupied states, respectively.
(b) The corresponding defect wavefunctions of the neutral \vnnb. These real space wavefunctions are visualized by reddish and greenish lobes representing the isosurface of the wavefunction at $+0.0002$~1/\AA$^3$ and $-0.0002$~1/\AA$^3$ values, respectively. For the sake of clarity, the atoms far from the defect are not shown.
(c) HSE Kohn-Sham defect levels of \vnnb~defect in the excited states of ${}^2A_1$, ${}^2B_2'$, and ${}^4A_1$.
(d) The permitted dipole transitions; thick red arrows refer to brighter axial transitions while orange solid arrows indicate the optical transitions induced by out-of-plane dipole moment. Possible non-radiative transitions induced by spin-orbit (blue) and spin-spin (cyan) interactions are shown by curly dashed lines. Optically active energy levels are highlighted by the green square.}
\label{fig:vnnb}
\end{figure*}

The geometry of this defect is shown in Fig.~\ref{fig:scheme}. As supported by previous \abinit simulations it has $C_{2v}$ point group symmetry~\cite{Tran2016a}. Our hybrid DFT calculations also support this conclusion [see Fig.~\ref{fig:vnnb}(b)]. Here we apply the molecular orbital technique to build up the total wavefunctions and hence our subsequent theory. The identification of defect energy levels in the band-gap as well as the localization of corresponding orbitals allows us to analyze them via `defect molecular diagram'.

Because of its hexagonal structure of the lattice the $\sigma$-dangling bonds around the vacancy are $sp^2$ orbitals, which result from hybridization of $2s$, $2p_x$, and $2p_y$ orbitals, hence they lie mostly on the plane of the layer. Meanwhile, the $2p_z$ orbitals perpendicular to the plane provide $\pi$-dangling bonds in the monolayer case. This is also valid for a multilayer membrane, as the inter-layer bonds are of weak Van der Waals nature. Therefore, the atoms in the neighboring layers do not affect significantly the defect dynamics. This has been verified by the \abinit computations~\cite{Tran2016a} where the electronic structure of point defects in mono- and three-layer h-BN membranes show negligible discrepancies. The $x$-axis is the symmetry axis of the \vnnb~defect which points from the vacancy to the nitrogen atom [Fig.~\ref{fig:scheme}]. The atomic dangling bonds are named after their variety ($\sigma$ or $\pi$) and the atom of origin: $\{\sigma_N, \sigma_{B_1}, \sigma_{B_2}, \pi_N, \pi_{B_1}, \pi_{B_2}\}$.
Construction of the symmetry-adapted molecular orbitals (MOs) facilitates our further analyses. They provide the basis functions that diagonalize the attractive Coulomb Hamiltonian of the defect. These MOs are linear combinations of the set of atomic orbitals listed above, which are the bases for the irreducible representations of the defect point group. One finds them by applying the projection method~\cite{Cornwell1997}. The MOs in the energy order from lowest to highest are~\cite{Abdi2017, Cheng2017}:
	$a_1^{(1)} = \alpha\sigma_N +\frac{\beta}{\sqrt{2}}(\sigma_{B_1} +\sigma_{B_2})$,
	$a_1^{(2)} = \beta\sigma_N +\frac{\alpha}{\sqrt{2}}(\sigma_{B_1} +\sigma_{B_2})$, 
	$b_2^{(1)} = \alpha'\pi_N +\frac{\beta'}{\sqrt{2}}(\pi_{B_1} +\pi_{B_2})$,
	$b_2^{(2)} = \beta'\pi_N +\frac{\alpha'}{\sqrt{2}}(\pi_{B_1} +\pi_{B_2})$,
	$b_1 = \frac{1}{\sqrt{2}}(\sigma_{B_1} -\sigma_{B_2})$,
	$a_2 = \frac{1}{\sqrt{2}}(\pi_{B_1} -\pi_{B_2})$,
where $\alpha$, $\beta$, $\alpha'$, and $\beta'$ coefficients are the overlap integrals and $\abs{\alpha}^2 +\abs{\beta}^2 = 1$ and $\abs{\alpha'}^2 +\abs{\beta'}^2 = 1$. We have named the single-electron orbitals after the symmetry of the irreducible representation they transform like.

In the ground state, the five defect electrons fill two orbitals and half-occupy the third. The \abinit calculations show that the fully occupied $a_1^{(1)}$ orbital lies deep in the valence band. Three of these orbitals with three occupant electrons are placed well-within the band-gap. Meanwhile, two unoccupied orbitals are located in the conduction band. The electronic configuration is sketched in Fig.~\ref{fig:vnnb}. The energy spacing of the orbitals is such that the effect of those located in valence and conduction bands can be safely neglected. We thus only consider the band-gap orbitals and electrons in our study.
Hence, the ground state is $[a_1]^2[b_2]^1[b_2']^0$ and a few possible excited states are $[a_1]^1[b_2]^2[b_2']^0$, $[a_1]^2[b_2]^0[b_2']^1$, and $[a_1]^1[b_2]^1[b_2']^1$. Here, we have adopted a shorthand for orbitals $b_2 \equiv b_2^{(1)}$ and $b_2' \equiv b_2^{(2)}$ and this convention will be used in our following analysis. The superscripts indicate the number of electrons occupying each orbital. In our DFT calculations the spin-polarized defect levels appear in the fundamental band-gap of h-BN. The occupation [Fig.~\ref{fig:vnnb}(a)] and symmetry [Fig.~\ref{fig:vnnb}(b)] of these wavefunctions fully supports the group theory analysis. The empty $b_2'$ spin-polarized levels lie in the gap, hence, intra-defect optical transitions are viable.

\subsection{Multi-electron states}
The multi-electron excited states calculated by our \abinit method are presented in Fig.~\ref{fig:vnnb}(c). The schematic energy levels with possible dipole induced and non-radiative transitions are also summarized in Fig.~\ref{fig:vnnb}(d). We combine the tensor products of $a_1$, $b_2$, and $b_2'$ states with the total spin eigenstates to construct the basis set, from which the multi-particle states compatible with $\bar C_{2v}$ can be calculated.
Here, the ground state and the two first excited states are constructed by a filled and a half-occupied orbital. Therefore, their multi-electron spin state can only take antisymmetric doublet state and all other combinations are rejected by Pauli's exclusion principle. For the third excited state, where three electrons occupy three different singlet orbitals, both doublet (symmetric and antisymmetric) and quartet spin states are possible.
The total wavefunction of these four electronic configurations in terms of Slater determinants are summarized in Table~\ref{tab:vnnb_khonsa}. (Note that the primes in ${}^2B_2'$, ${}^2A_1'$, and ${}^2A_1''$ do not correspond to other irreducible representations and are to distinguish the states). The order of energy levels in the third excited state is estimated by Hund rules: The state with highest multiplicity lies at lower energies. Therefore, the quartet comes first, then the doublet with antisymmetric space wavefunction---which minimizes the Coulomb repulsion of electrons.
As we will see below, the four-fold degeneracy of the quartet states is lifted and reduced into two two-fold degeneracies by the electronic spin-spin dipole interaction.

\renewcommand{\arraystretch}{1.3}
\begin{table}[tb]
\caption{\label{tab:vnnb_khonsa} The configuration and spin-orbit total wavefunctions of neutral \vnnb~in terms of superpositions of the Slater determinants. The prime sign in $\mathrm{d}'$ stands for symmetric nature of the doublet spin wavefunction. The line over orbitals in the Slater states indicate the spin-down state of the electron in that orbital.}
\begin{ruledtabular}
\begin{tabular}{lclc}
configuration & ${}^{2S+1}\Gamma_o$ & \multicolumn{1}{c}{Slater linear superpositions} & label\\
\hline
$[a_1]^2[b_2]^1[b_2']^0$ & $\prescript{2}{}{B}_2$ & $\ket{a_1\overline{a}_1b_2}, \ket{a_1\overline{a}_1\overline{b}_2}$ & $\GSD$ \vspace{1mm}\\
$[a_1]^1[b_2]^2[b_2']^0$ & $\prescript{2}{}{A}_1$ & $\ket{a_1b_2\overline{b}_2}, \ket{\overline{a}_1b_2\overline{b}_2}$ & $\FED$ \vspace{1mm}\\
$[a_1]^2[b_2]^0[b_2']^1$ & $\prescript{2}{}{B}_2'$ & $\ket{a_1\overline{a}_1b_2'}, \ket{a_1\overline{a}_1\overline{b}_2'}$ & $\SED$ \vspace{1mm}\\
$[a_1]^1[b_2]^1[b_2']^1$ & $\prescript{4}{}{A}_1$ & $\ket{\overline{a}_1b_2b_2'}+\ket{a_1\overline{b}_2b_2'}+\ket{a_1b_2\overline{b}_2'}$, & $\mathscr{A}^{3,\rm q}_{+\sfrac{1}{2}}$ \\
  &   & $\ket{\overline{a}_1\overline{b}_2b_2'}+\ket{\overline{a}_1b_2\overline{b}_2'}+\ket{a_1\overline{b}_2\overline{b}_2'}$ & $\mathscr{A}^{3,\rm q}_{-\sfrac{1}{2}}$ \vspace{1mm}\\
  &   & $\ket{a_1b_2b_2'}, \ket{\overline{a}_1\overline{b}_2\overline{b}_2'}$ & $\TEQu$ \vspace{1mm}\\
  & ${}^2A_1'$ & $\ket{\overline{a}_1b_2b_2'}+\ket{a_1\overline{b}_2b_2'}-2\ket{a_1b_2\overline{b}_2'}$, & $\mathscr{A}^{3,\rm d'}_{+\sfrac{1}{2}}$ \\
  &   & $\ket{\overline{a}_1\overline{b}_2b_2'}+\ket{\overline{a}_1b_2\overline{b}_2'}-2\ket{a_1\overline{b}_2\overline{b}_2'}$ & $\mathscr{A}^{3,\rm d'}_{-\sfrac{1}{2}}$ \vspace{1mm}\\
  & ${}^2A_1''$ & $\ket{a_1\overline{b}_2b_2'}-\ket{\overline{a}_1b_2b_2'}$, & $\mathscr{A}^{3,\rm d}_{+\sfrac{1}{2}}$ \\
  &   & $\ket{a_1\overline{b}_2\overline{b}_2'}-\ket{\overline{a}_1b_2\overline{b}_2'}$ & $\mathscr{A}^{3,\rm d}_{-\sfrac{1}{2}}$
\end{tabular}
\end{ruledtabular}
\end{table}
\renewcommand{\arraystretch}{1.0}

We calculated the lowest excitation energies that may be directly compared to experimental data, and reveal the position of the quartet level with respect to the doublet excited states' levels. Since the quartet level may play a role in the intersystem crossing (ISC) processes, thus the knowledge about the order of these levels is of high importance. We created the excited states in the $\Delta$SCF procedure as obtained in the group theory analysis, and the spin-polarized levels and their occupation are shown in Fig.~\ref{fig:vnnb}(c). The corresponding excitation energies are listed in Table~\ref{tab:VN_NB-energies}.
We demonstrate for this defect that the popular semi-local PBE DFT functional~\cite{Perdew1996} strongly underestimate the zero-phonon-line energies, whereas the HSE provides reliable energies. Indeed, the calculated first zero-phonon-line excitation energy is close to the detected one~\cite{Tran2016a}. In the previous studies, the calculated vertical excitation energy of the defect were inaccurately compared to the measured zero-phonon-line energy at PBE DFT level.
Our calculations show that the error in the calculated ZPL by PBE functional and the neglect of relaxation energy almost cancels each other. Overall, the calculated HSE ZPL energy of charge neutral \vnnb~indeed shows good correspondence with the signature of the single-photon emitter in h-BN. Interestingly, the calculated quartet $^{4}A_1$ level is between the two doublet excited states' levels. Although, the calculated gaps between these levels are relatively large, ultraviolet or optical two-photon excitations may lead to ISC processes.

Furthermore, it is worth mentioning that ${}^2B_2$ and ${}^2B_2'$ have the same symmetry and thus the Coulomb correlation effects can mix them up. That is, the states that diagonalize the Coulomb Hamiltonian are $\ket\GSD +\kappa\ket\SED$ and $\kappa\ket\GSD +\ket\SED$. The large energy difference between the two states $3.65$~eV, on the other hand, is expected to reduce the degree of mixing such that $\kappa \approx 0$. Hence, for our current study the mixing will be neglected and the multi-particle states listed in Table~\ref{tab:vnnb_khonsa} are assumed to be reasonably valid. This effect, however, can slightly modify the dipole allowed transitions of the system and thus the excitation-emission dynamics of the system as we will discuss in below.

\subsection{Spin interactions}
We now study the effect of the spin interactions on electronic structure and dynamical properties of the defect by applying time-independent perturbation theory.
As we will find out, the spin-orbit and spin-spin interactions do not lift the Kramers degeneracy of the states. However, they can provide non-radiative transitions. In a short notation, the spin-orbit interaction Hamiltonian is $H_{\rm so}=\sum_j\sum_\alpha \ell^\alpha_j s^\alpha_j$, where $j=1,2,3$ counts the particle numbers, while $\alpha$ addresses the irreducible representations by which the orbital and spin angular momenta transform \cite{Lenef1996, Marian2001}.
Here, $\bm\ell_j = (1/2m^2c^2)\bm\nabla V(\bm{r}_j)\times\bm{p}_j$ with $V(\bm{r})$ the nuclei Coulomb potential and $\bm{s}_j$ are the spin and angular momenta vectors of the $j$-th electron, respectively.
The group theory allows us to predict potentially nonzero matrix elements of a Hamiltonian. In general, the matrix element $\bra\psi O \ket\phi$ vanishes if $\Gamma(\psi)\otimes\Gamma(O)\otimes\Gamma(\phi) \not\supset \Gamma_1$, where $\Gamma(X)$ is the irreducible representation of operator or wavefunction $X$ and $\Gamma_1$ is the totally symmetric irreducible representation. 
One thus concludes from $C_{2v}$ character table that only $\ell_y$ component of the orbital angular momentum could give nonzero values. (The angular momenta transform like axial vectors and have no $A_1$ component). This simplifies the spin-orbit Hamiltonian to $H_{\rm so}=\sum_j\ell_{y,_j}s_{y,_j}$.

\renewcommand{\arraystretch}{1.2}
\begin{table}[tb]
\caption{\label{tab:VN_NB-energies} Optical transition energies of \vnnb~defect as calculated by HSE $\Delta$SCF method in units of electronvolt.}
\begin{ruledtabular}
\begin{tabular}{lccl}
 & $^{2}A_1$ & $^{2}B_2'$ & $^{4}A_1$ \\
\hline
Vertical absorption energy & 2.53 & 4.02 & 3.17 \\
     Relaxation energy     & 0.48 & 0.37 & 0.34 \\
     Zero phonon line      & 2.05 & 3.65 & 2.83\footnote{Zeroth order dipole transitions are forbidden.} \\
     Zero phonon line, PBE     & 1.40 & 2.98 & 2.54\textsuperscript{a} \\
\end{tabular}
\end{ruledtabular}
\end{table}
\renewcommand{\arraystretch}{1.0}
%

Another effect of spin is the direct magnetic dipole moment interaction between the electrons and nuclei. We neglect the interaction of nuclear spin with the electronic spins, which introduce hyperfine structures to the system and only focus on the electronic system. The spin-spin interaction is
$H_{\rm ss}=\frac{\mu_0}{4\pi}\gamma_e^2\hbar^2
			\sum_{j>k}\frac{1}{r_{jk}^3}\big[\bm{s}_j.\bm{s}_k -3(\bm{s}_j.\bm{\widehat r}_{jk})(\bm{s}_k.\bm{\widehat r}_{jk})\big]$,
where $\bm{r}_{jk}=\bm{r}_j-\bm{r}_k$ is the displacement vector between electron $j$ and electron $k$, and $\gamma_e$ is the gyromagnetic ratio of electron.
This Hamiltonian can be rewritten as
\begin{equation}
H_{\rm ss}=\frac{\mu_0}{4\pi}\gamma_e^2\hbar^2\sum_{j > k}\sum_{\alpha} \widehat{S}_{jk}^\alpha \widehat D^\alpha_{jk},
\label{hss}
\end{equation}
where $\widehat D^{\alpha}_{jk}$ are the symmetry-adapted spin-spin second rank tensor components, e.g. $\widehat D^{B_2}_{jk}=(1/2r_{jk}^5)(x_{jk}z_{jk}+z_{jk}x_{jk})$~\cite{Marian2001}. We have also expressed spin vector of the two electrons in the dyadic form $\widehat S \equiv \bm s\bm s$. 
Only $\widehat D^{A_1} = \{\widehat D_{xx}, \widehat D_{yy}, \widehat D_{zz}\}$ and $\widehat D^{B_2} = (\widehat D_{xz} +\widehat D_{zx})/2$ components can give nonzero values and the rest have no contribution.
The $\widehat D^{A_1}$ components simply impose energy shifts on all states. Most importantly, the interaction lifts degeneracy of the quartet state ${}^4A_1$ such that $\TEQu$ assume higher energies than $\TEQd$.
Another noticeable effect of this interaction is mixing states with different spin projections $m_S$ inducing non-radiative transitions channels [Fig.~\ref{fig:vnnb}(d)]. For the explicit form of spin-orbit Hamiltonian alongside the spin-spin interaction we refer the reader to the supporting information.

\subsection{Selection rules}
Next we study possible optical transitions in the system. In the lowest order such transitions can happen via the electric dipole interaction: $H_{\rm dp}=\sum_j\sum_\alpha d^\alpha_jE^\alpha_j$, where $\bm{d}=-e(x,y,z)$ is the dipole moment of the electron and $\bm{E}$ is the electric field vector.
Any polar vector (including the dipole moment) in $C_{2v}$ transforms like $(B_1,B_2,A_1)$. In calculating the matrix elements, spinors are just spectator component of the wavefunction. Hence, the orbital symmetries and spin overlaps determine the allowed transitions.
Generally, since the orbitals have either $A_1$ or $B_2$ point-group irreducible symmetry, the allowed transitions are either induced by the axial in-plane component of the dipole moment $d_x$ (between orbitals with the same symmetry) or the out-of-plane component $d_z$. The transition rates are proportional to the quantities $\bra{b_2}d_x\ket{b_2'}E_x$, $\bra{a_1}d_z\ket{b_2}E_z$, and $\bra{a_1}d_z\ket{b_2'}E_z$.
According to orthogonality of the orbitals and spin states, the allowed optical dipole transitions are those sketched in Fig.~\ref{fig:vnnb}(d).

%
%
\section{Negatively Charged \vb}
The geometry of a boron vacancy is shown in Fig.~\ref{fig:scheme}. The negatively charged \vb~has $D_{3h}$ point group symmetry~\cite{Huang2012a}. Similar to \vnnb\ defect discussed in the previous section there are three hybrid $sp^2$ dangling-bonds and three $2p_z$ orbitals. In this case the atoms are identical and hence are their dangling-bonds: $\{\sigma_1, \sigma_2, \sigma_3, \pi_1, \pi_2, \pi_3 \}$. The single-electron MOs can be obtained by applying the projection operators produced from the character table of the $D_{3h}$ on these dangling-bonds.
The equivalence representation for the structure of \vb~defect suggests that orbitals with $A_1'$, $A_2''$, $E'$, and $E''$ should be present. The MOs are then obtained:
	$a'_1 = \frac{1}{\sqrt{3}}(\sigma_1 +\sigma_2 +\sigma_3)$,
	$e''_x = \frac{1}{\sqrt{6}}(2\pi_1 -\pi_2 -\pi_3)$,
	$e''_y = \frac{1}{\sqrt{2}}(\pi_2 -\pi_3)$,
	$a''_2 = \frac{1}{\sqrt{3}}(\pi_1 +\pi_2 +\pi_3)$,
	$e'_x = \frac{1}{\sqrt{6}}(2\sigma_1 -\sigma_2 -\sigma_3)$,
	$e'_y = \frac{1}{\sqrt{2}}(\sigma_2 -\sigma_3)$,
where the $e_x'$ and $e_y'$ single-electron orbitals are degenerate as well as $e_x''$ and $e_y''$. In a neutral defect, every atom shares three electrons with the other ions in the defect. Hence, for the negatively charged case there is a total number of ten dynamical electrons. 

\begin{figure*}[tb]
\includegraphics[width=0.7\textwidth]{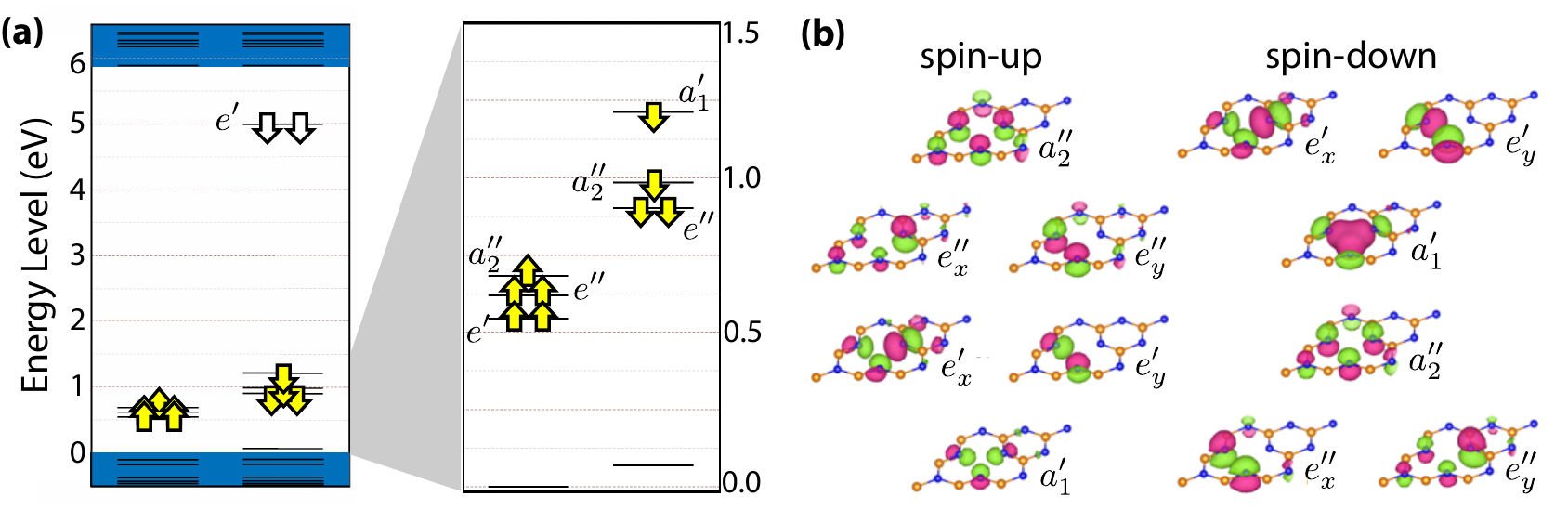}
\caption{%
(a) The ground state HSE Kohn-Sham defect levels of the negatively charged boron vacancy in the fundamental band gap of h-BN. The spin-up level of $a_1'$ falls in the valence band. The yellow and white arrows represent the occupied and unoccupied states, respectively.
(b) The defect wavefunctions corresponding to the molecular orbitals. These real space wavefunctions are visualized by the reddish and greenish lobes representing the isosurface of the wavefunction at $+0.0002$~1/\AA$^3$ and $-0.0002$~1/\AA$^3$ values, respectively.}
\label{fig:vb}
\end{figure*}

Our HSE DFT spin-polarized calculation on the negatively charged \vb\ shows a complicated electronic structure [see Fig.~\ref{fig:vb}(a)]. The spin polarization of the defect states is large. The spin-up level of the $a_1'$ state fall in the valence band, thus only 9 electrons are visible with 5 spin-up and 4 spin-down electrons in the fundamental gap. By taking into account the spin-up $a_1'$ level in the valence band, this results in an $S=1$ ground state, as predicted by our group theory analysis (see below). The double degenerate $e'$ and $e''$ levels lie very close to each other, and also to the $a_2''$ level. This implies relatively small difference in the optical excitation energies of the corresponding many-body excited states.

%

\subsection{Multi-electron states}
The `physical' total wavefunctions are determined by direct multiplication of the irreducible spatial and spin wavefunctions. Of course, at the end an anti-symmetrization is required by writing the wavefunctions in terms of linear combinations of Slater determinants. Since the number of electrons contributing in the defect dynamics is rather high in this case, it is more convenient to deal with `hole's (lack of electrons) instead~\cite{Maze2011, Doherty2011}.

\subsubsection*{Ground state}
In the ground state, two holes are in the $e'$ degenerate orbitals.
Therefore, spatial part of the wavefunction transforms like $E'\otimes E' = A_1'\oplus A_2'\oplus E'$ and the corresponding functions are found by projection technique.
The spin part on the other hand has $A_1'\oplus A_2'\oplus E''$ symmetries, which are the two-fermion singlet $\chi^{A_1'}$ and triplet $\{\chi^{A_2'},\chi^{E_x''},\chi^{E_y''}\}$ states.
To construct the total wavefunctions one systemically multiplies the spatial states by the spinors, the outcomes are indeed superpositions of Slater determinants and only the physical states will survive. That is, the states that do not violate Pauli's exclusion principle.
One then finds that the ground state manifold is composed of $\{{}^3A_2',{}^1E_x',{}^1E_y',{}^1A_1'\}$.

The electron-electron repulsion Coulomb energy can be easily computed here. The Hund rules imply that the triplet state ${}^3A_2'$ lies at the lowest level and gives the ground state. While the $E'$ and $A_1'$ singlet states place in higher energies with equal energy spacings. This, in fact, is concluded from the symmetry observation that the Coulomb potential is a scalar and transforms as $A_1'$, and therefore, the energy expectation value for both $E'$ states should be the same. Then, the explicit calculations show that the Coulomb energy difference of ${}^3A_2'$ and ${}^1E_y'$ equals the energy difference of ${}^1A_1'$ and ${}^1E_x'$ and it exactly is given by Coulomb exchange energy.

\subsubsection*{Excited state}
The spin symmetry and possible spin states of the defect excited states are the same as the ground state.
The first excited state, as inferred from our HSE Kohn-Sham computations presented in Fig.~\ref{fig:vb}(a), is attained when an electron from the state $a_1'$ climbs to the $e_x'$ or $e_y'$ states. In this case, the only irreducible symmetry of the spatial wavefunction is $A_1'\otimes E' = E'$ and hence the wavefunctions are simply $\frac{1}{\sqrt 2}(a_1'e' \pm e' a_1')$. They thus can assume both triplet and singlet spin components: $\{{}^3E_x',{}^3E_y',{}^1E_x',{}^1E_y'\}$.
Meanwhile, with a very small energy difference, it is also probable to have the electron transferred from $a_2''$ to $e'$ doublet in the spin-down channel. The spatial wavefunctions then will have $A_2''\otimes E' = E''$ irreducible symmetries. The explicit form of the wavefunctions, in analogy to the first excited state, are $\frac{1}{\sqrt 2}(a_2''e' \pm e'a_2'')$. This shapes the manifold $\{{}^3E_{x,a}',{}^3E_{y,a}',{}^1E_{x,a}',{}^1E_{y,a}''\}$.
The spatial part, for the next possible excited state, is composed of $E'\otimes E'' = A_1'' \oplus A_2'' \oplus E''$ symmetries. A triplet and singlet state is assigned to every symmetric and antisymmetric function, respectively.
Their energy degeneracy is lifted by introducing the Coulomb repulsion energies and according to Hund's rules the spin singlet states with lowest orbital momenta attain the highest energies whereas the triplet states with largest angular momenta (here the state with $E''$ symmetry) lie at the lowest energy forming up the triplets $\{{}^3E_{x,\pi}'',{}^3E_{y,\pi}'',{}^3A_2'',{}^3A_1''\}$ and singlets $\{{}^1E_{x,\pi}'',{}^1E_{y,\pi}'',{}^1A_2'',{}^1A_1''\}$.

Our HSE Kohn-Sham levels indicate that, indeed, the ${}^3E_a''$ excitation energy (promoting an electron from the $a_2''$ orbital to the $e'$ orbitals) is lower in energy than the ${}^3E'$ in $a_1'e'$ configuration and ${}^3\Gamma''$ excitations ($\Gamma ={A_1, A_2, E_\pi}$) which can be constructed by promoting an electron from the $e''$ to the $e'$ state. The calculated energy differences with respect to the ground state ${}^3A_2'$ are 2.13~eV and 1.92~eV in $D_{3h}$ symmetry for the ${}^3E_a''$ and ${}^3E'$ states. respectively. The latter, as we will discuss below, has a dipole allowed transition to the ground state.
Basically, ${}^3E_a''$ and ${}^3E'$ states are Jahn-Teller systems. In the Born-Oppenheimer approximation (that we inherently apply in our DFT simulations) we find that in the $C_{2v}$ configuration they assume lower energies than the high symmetry $D_{3h}$ configuration. This energy difference is the so-called Jahn-Teller energy of these states. The effect is such that the arrangement of ${}^3E_a''$ and ${}^3E'$ states is reshuffled.
In experiment, we expect that the calculated ZPL energy of the $C_{2v}$ configuration can be observed for ${}^3E'$.
However, the large phonon energies of the h-BN lattice can result in a large electron-phonon coupling which results in a dynamic Jahn-Teller system showing a high symmetry in the experiments (average of the three $C_{2v}$ configurations), at least, at room temperature. Thus, the selection rules worked out for the high symmetry are expected to be valid at ambient conditions. It is beyond the scope of our study to quantitatively estimate the electron-phonon coupling here, and explore the fine structure at cryogenic temperatures.

Now, we turn to the estimation of the other excitation energies. The ${}^3\Gamma''$ excited states in $e'e''$ configuration are highly complicated: they can be described by Slater multi-determinants that cannot be accurately calculated within $\Delta$SCF procedure. In addition, strong electron-phonon interactions can also couple these states. On the other hand, a crude estimate for the average energy of these states can be yielded from the $\Delta$SCF procedure, i.e., promoting an electron from the $e''$ to the $e'$ state. The calculated total energy of this state is $\approx 0.7$~eV larger in energy than the lowest excitation energy. This result implies that the ${}^3E_a''$ and ${}^3E'$ states is below the levels of ${}^3A_1''$, ${}^3A_2''$, and ${}^3E_\pi''$ states [see Table~\ref{tab:VB-energies}].

\renewcommand{\arraystretch}{1.2}
\begin{table}[tb]
\caption{\label{tab:VB-energies} Energy levels of negatively charged \vb\ excited states with respect to the ground state as calculated by HSE $\Delta$SCF method in units of electronvolt.}
\begin{ruledtabular}
\begin{tabular}{cccl}
symmetry & $^{3}E_a''$ & $^{3}E'$ & $^{3}E_\pi''$ \\
\hline
	$D_{3h}$ & 2.13 & 1.92 & 2.62 \\
    $C_{2v}$ & 1.71 & 1.77 & 2.47 \\
\end{tabular}
\end{ruledtabular}
\end{table}
\renewcommand{\arraystretch}{1.0}
%

\subsubsection{Spin interactions}
The spin and orbital angular momenta components transform like $\{E''_x,E''_y,A'_2\}$. Regarding orbital part; since none of the holes are in molecular orbital with $E''$ symmetry in the ground state, the interaction should only come from the axial component of the orbital angular momentum. In the excited state, matrix elements like $\bra{e'_x}\ell_x\ket{e'_y}$ could still contribute in the dynamics as they are predicted by their symmetry to be different from zero as they contain the $A_1'$ irreducible representation. However, the purely imaginary character of the angular momenta operator demands that such matrix elements must vanish~\cite{Lenef1996}. This reduces the spin-orbit Hamiltonian into $H_{\rm so} = \ell_{z,1}s_{z,1} +\ell_{z,2}s_{z,2}$.
The most important effect of spin-orbit interaction in the excited states manifolds is lifting the degeneracy of ${}^3E_a''$, ${}^3E'$, and ${}^3E_\pi''$ states. Such that the states with maximum spin projection absolute value $\abs{m_S} = 1$ get mixed. The energy of states with $m_S=0$ remains intact by the spin-orbit interaction. Instead, these states are coupled to the singlet states and may result-in irradiative transition channels.

\begin{figure}[tb]
\includegraphics[width=\columnwidth]{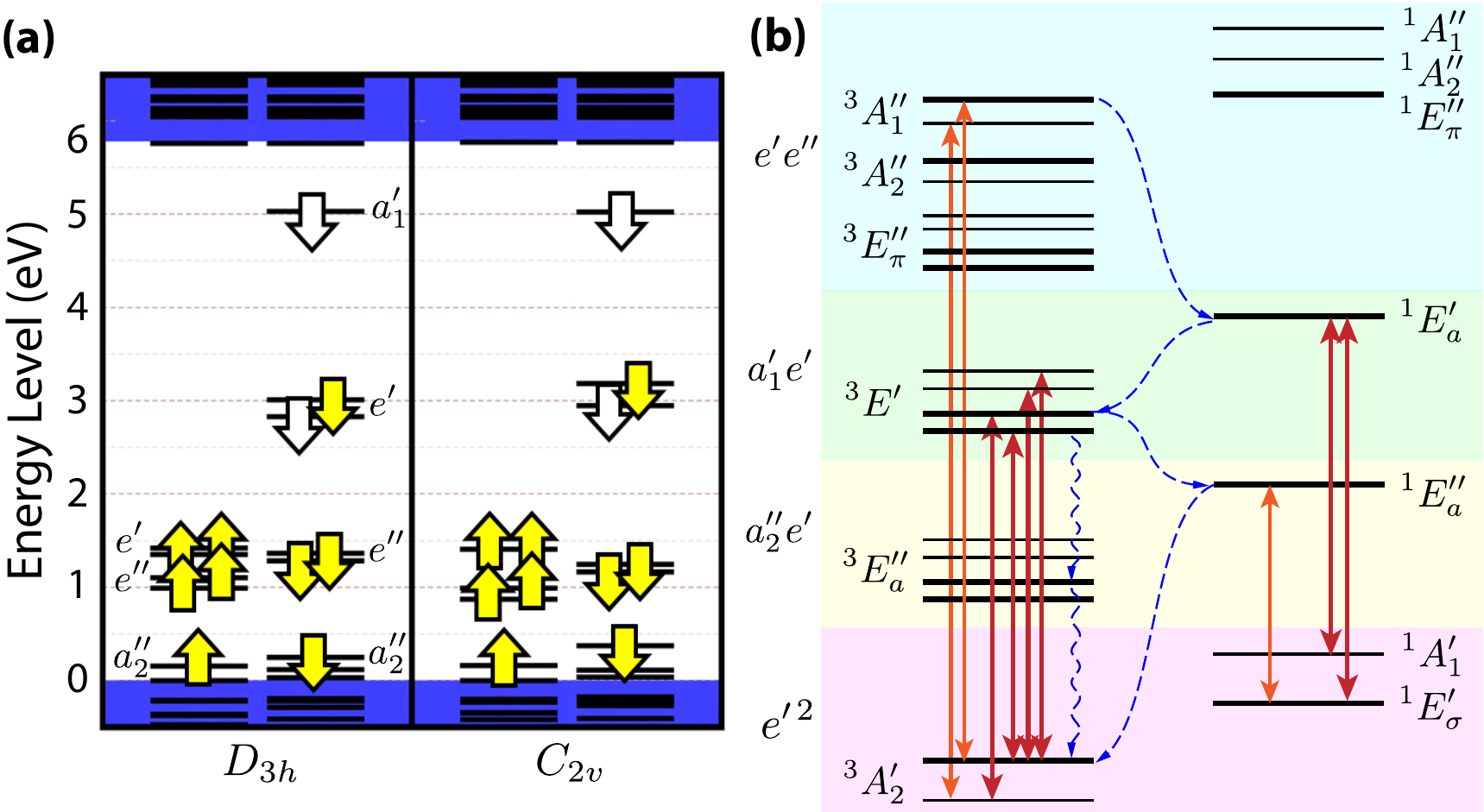}
\caption{Energy levels of negatively charged \vb~defect in h-BN:
(a) HSE Kohn-Sham levels for the ${}^3E'$ excited state with $D_{3h}$ and $C_{2v}$ symmetries.
(b) Electronic and spin configuration of the defect. The triplet (left column) and singlet (right column) states are separated for clarity. The radiative transitions are marked by arrows in different colors for in-plane (red) and out-of-plane (orange) dipole moments. The possible spin-orbit induced non-radiative transitions that will lead to the optical spin polarization of the ground state are also shown by blue dashed lines. Energy differences are not to the scale.}
\label{fig:vbtrans}
\end{figure}

Finally, this interaction makes connections between different electronic states. That is, inter-electronic admixture of the singlet and triplet spin states as well as mixing the states within triplet and singlet spin states. Therefore, allowing for electronic or vibronic induced transitions that do not conserve the spin projection. Such transition might be non-radiative, too. In a nutshell, the spin-orbit interaction mixes states with the same total wavefunction symmetry and different $\abs{m_S}$ values. We refer the reader to Fig.~\ref{fig:vbtrans}(b) for a qualitative picture of such state mixing induced transitions. The explicit form of the spin-orbit interaction matrix elements are too cumbersome to be reported here.

The effect of spin-spin interaction can be summarized as follows. The spin degeneracy in triplet states is lifted between states with different $|m_S|$: The axial component $\widehat D^{A_1'}$ imposes energy differences between $m_S=0$ and $m_S=\pm 1$ states.
Another important effect is further lifting the orbital degeneracies in ${}^3E_a''$, ${}^3E'$, and ${}^3E_\pi''$. The energy of all singlet states is increased more that their triplet counterpart, which cannot influence the level ordering because the spin interactions are much weaker than Coulomb repulsions and thus do not change order of the states. The electronic structure of negatively charged \vb\ states associated with spin-orbit and spin-spin induced splittings are shown in Fig.~\ref{fig:vbtrans}(b).
The spin-spin interaction of the electrons also induces electronic and spintronic mixings, which for the sake of clarity and simplicity are not reported here.

%

\subsection{Selection rules}
The electrical dipole moment $\bm{d}~(=-e\bm{r}$) components transform like $(E'_x,E'_y,A''_2)$ in $D_{3h}$ point group symmetry. In the zeroth order of spin interaction perturbation, the spin component of the wavefunctions imply that only singlet-singlet and triplet-triplet transitions are permitted. It is noteworthy that when the first order perturbation of the spin-orbit and spin-spin interactions are taken into account the spin flip transitions are also possible.
From symmetry observations, one already finds that the first excited state is a dark state, while non-axial components of the dipole moment can cause excitations and emissions to and from the second excited state. The bright triplet ${}^3E'$ states can relax into the ground triplet ${}^3A_2'$ via different competing processes; radiative, ISC, and direct non-radiative.
Due to the spin-orbit mixing in ${}^3E'$ manifold the emitted photons stemming from transitions from these states to the ground state will adopt different polarizations. For states with $m_S=\pm1$ the emitted photons should have circular polarizations ${}^3A_2' \xleftrightarrow{x\pm iy} {}^3E'$, while the $m_S=0$ states are connected via linearly polarized optical photons $^{3}A_2' \xleftrightarrow{\{x,y\}} {}^3E'$.
The other possible transitions are those to the third excited state. These transitions that are only induced by the axial dipole moment are ${}^3A_2' \xleftrightarrow{z} {}^3A_1''$ in the triplet manifold. The singlet states are expected to assume much higher energies, and therefore, optically inaccessible.
These radiative transitions are sketched in the energy level diagram in Fig.~\ref{fig:vbtrans}(b) alongside the possible spin-orbit interaction induced non-radiative transitions, which encompass optical spin polarization channels.
The situation in the singlet channel is similar for the second and third excited states. However, a $d_z$-induced transition is possible between ${}^1E_\sigma'$ in the ground state manifold and ${}^1E_a''$ of the first excited state [Fig.~\ref{fig:vbtrans}(b)].

%
%
\section{Discussion}
Here we discuss our results in the context of the main experimental observations. The h-BN emitters are mainly divided into two classes based on their line shape and excitation/emission polarization pattern: (I) Those with an asymmetric and broader line shape that have a matching in-plane excitation and emission polarizations. (II) The SPEs that possess a sharp and symmetric line shape with a mismatch between their excitation and emission polarization patterns. The emission from class-I SPEs are typically around $2.1$~eV, while class-II emitters on average assume energies about $1.8$~eV~\cite{Tran2016b, Jungwirth2016}. Emitters of both classes exhibit several shelving states and non-radiative relaxation processes that compete with the radiative transitions. Moreover, two-photon processes are reported in the excitation of some emitters~\cite{Schell2016}.
It is also noteworthy that the peaks in photoluminescence spectra of mono- and multi-layer samples have significantly different widths. Due to their vulnerability to environmental effects, the emitters on a mono-layer h-BN exhibit much broader linewidths. Strain and quantum confinement effects in h-BN flakes may also influence the optical properties of defects residing at the edge of the flakes. Our theoretical work, instead, is performed by neglecting environmental effects. Therefore, our following comparisons are made with the experimental data where such environmental effects are minimized and the material is still two-dimensional. That is, those related to the multi-layer samples.

\begin{figure}[tb]
\includegraphics[width=0.7\columnwidth]{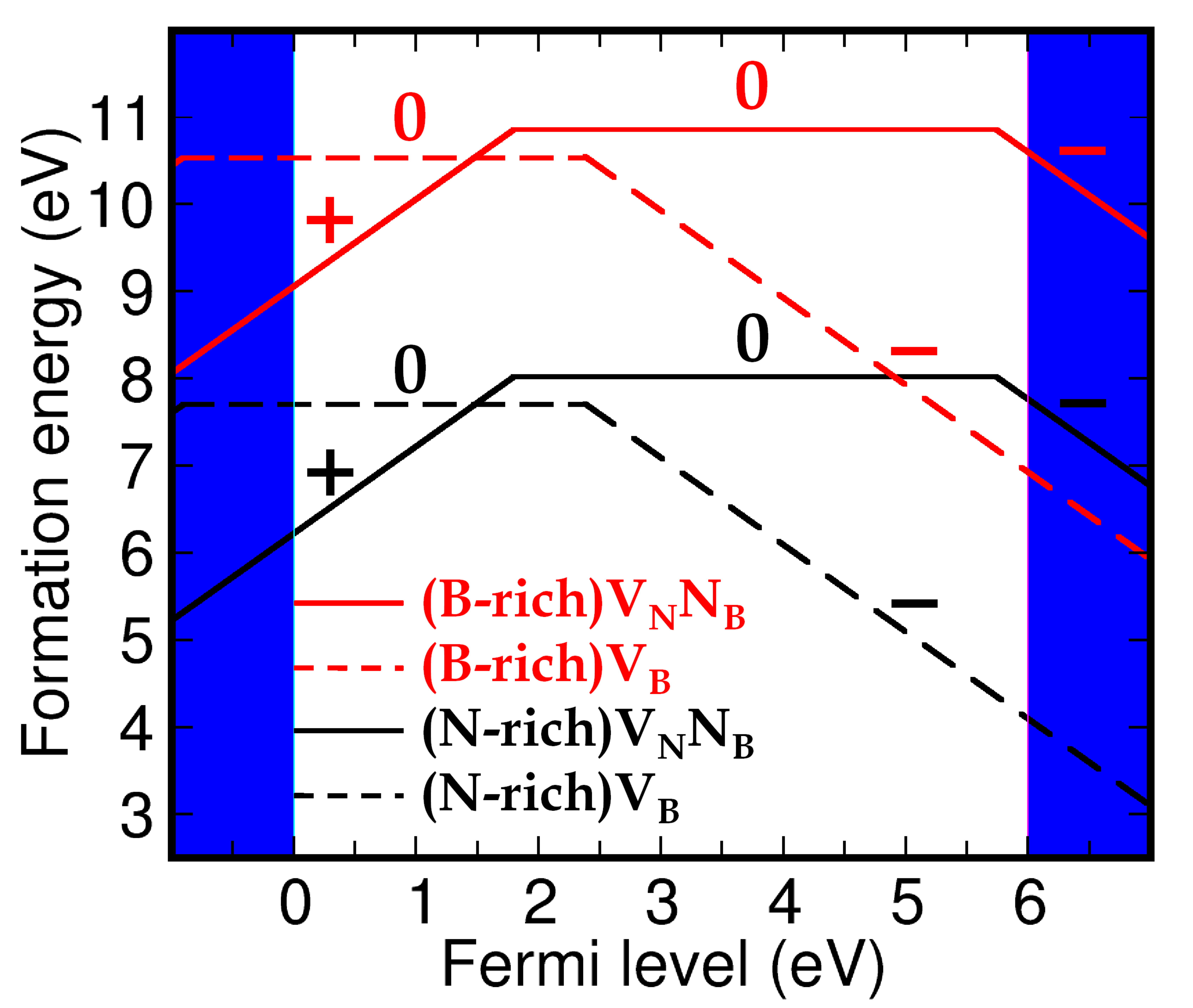}
\caption{Formation energy as a function of the Fermi level for defect \vb~(solid lines) and \vnnb~(dashed lines) in N-rich (black curve) and B-rich (red curve) growth conditions. The symbols indicate charge state of the defects.}
\label{fig:CTL}
\end{figure}

In particular, we consider the neutral \vnnb~defect and the negatively charged \vb~defect as qubit candidates. The \vb~and \vnnb~defects have the same composition in the compound h-BN material. The relative stability of these defects may depend on their charge state. We analyze this issue by calculating the defect formation energies ($E^q_f$) with charge state $q$, which is defined as
$E^q_f(\epsilon_F)=E^q_\text{tot}-E_\text{BN}+\mu_\text{B}+
q(\epsilon_\text{F}+E_\text{V})+E_\text{corr}$,
where $E^q_\text{tot}$ is the total energy of the charged defect system, $E_\text{BN}$ is the total energy of the pristine h-BN, $\mu_\text{B}$ is the chemical potential of boron, $\epsilon_F$ is the position of the Fermi-level with respect to the valence band maximum $E_\text{V}$, and $E_\text{corr}$ is the charge correction energy~\cite{Zhang1991,Wu2017}.
The boron chemical potential depends on the growth conditions. In nitrogen-rich conditions, the nitrogen atoms in h-BN are assumed to be in equilibrium with N$_2$ gas, therefore, $\mu_\text{N}$ equals half of the energy of a N$_2$ molecule ($\mu_\text{N}=\frac{1}{2}\mu_{\text{N}_2}$) and $\mu_\text{B}$ can be obtained from $\mu_\text{BN} = \mu_\text{B} + \mu_\text{N}$, where $\mu_\text{BN}$ is the energy of h-BN primitive cell.
The calculated HSE formation energies as a function of Fermi level are plotted in 
Fig.~\ref{fig:CTL}. The  ($+$) charge state is not stable for \vb~defect whereas its ($0/-$) 
acceptor level is at $E_\text{V}+2.39$~eV. For \vnnb~defect, the ($+/0$) level is at 
$E_\text{V}+1.79$~eV 
whereas the deep ($0/-$) acceptor level occurs only for high $\epsilon_{\rm F}$ values. That is, at 0.24~eV below the conduction band edge (HSE bandgap is 5.98~eV), which is fairly consistent with a very recent result~\cite{Wu2017}.

We find that the formation energies of these defects are high even at nitrogen-rich condition that is favorable for \vb-like defects. On the other hand, we note that the nitrogen chemical potential may vary significantly at realistic experimental conditions of N$_2$ gas such as partial pressure and temperature. This, however, affects the absolute values in the calculated formation energies but not their relative values. Thus, we rather focus on the relative stability of \vb~and \vnnb~defects. Our findings show that these defects exhibit a bistability: at low Fermi-level values (p-type conditions) the \vnnb~defect is stable whereas \vb~defect becomes stable only for $\epsilon_\text{F}>1.5$~eV. In the region of $1.9$~eV$<\epsilon_\text{F}<2.6$~eV the neutral \vnnb\ is almost as stable as the neutral \vb\ within $\approx0.2$~eV. The stability of negatively charged \vb\ becomes dominant for $\epsilon_\text{F}>2.6$~eV. We conclude that the neutral \vnnb~and negatively charged \vb~defects can indeed exist in h-BN and as discussed below can be the source of single photon emissions.

Next, we discuss the neutral \vnnb~defect as a quantum emitter in h-BN.
The 2D nature of layered h-BN samples demands that the \vnnb~defect axis of symmetry remain in the plane of the membrane. Yet the orientation of the defect symmetry axis is restricted to: $\vartheta = 0^\circ, 120^\circ, 240^\circ$, where $\vartheta$ is the angle between defect symmetry axis and the $x$-axis in the lab [Fig.~\ref{fig:scheme}].
A perpendicular optical beam irradiating a h-BN flake, thus, preferentially excites the in-plane dipole moment of the defect. 
The magnitude of the exciting dipole moment is given by $d_x\cos\vartheta$ for the three possible orientations. In zeroth order, the axial moment can only induced transitions between the ground state and the second excited state ${}^2B_2\leftrightarrow {}^2B_2'$. Given the typical $\approx 2.4$~eV excitation lasers used in the experiment, the ${}^2B_2'$ energy level is inaccessible via single-photon transitions according to our DFT simulations [Table~\ref{tab:VN_NB-energies}]. Therefore, the charge neutral \vnnb~defects do not follow (in the zeroth order) the reported polarization pattern; that the excitation and emission dipole moments are similarly oriented in the plane of flake~\cite{Tran2016a}.
Nevertheless, one notices that the out-of-plane dipole moment $d_z$ still can give rise to excitations and emissions at energies around $2.0$~eV due to ${}^2B_2 \leftrightarrow {}^2A_1$ coupling, see Table~\ref{tab:VN_NB-energies}. Moreover, two photon processes can excite the defect into its second excited state ${}^2B_2'$. This can either stem from the real excitation via ${}^2A_1$ state induced by the $d_z$ dipole moment or the direct nonlinear two-photon excitation facilitated by the in-plane $d_x$ dipole moment~\cite{Schell2016}. The `real' two-photon excitation of emitters in h-BN may occur by climbing the energy ladder thanks to the Coulomb mixing effect between ${}^2B_2$ and ${}^2B_2'$ states discussed above.
The defect excited to ${}^2B_2'$ will face several competing routes to relax down to the ground state: The non-radiative and radiative shelving processes or a mixture of them [Fig.~\ref{fig:vnnb}(d)]. 
The existence of the shelving state ${}^4A_1$ predicted by our DFT and group theory study can be experimentally confirmed by applying excitation photons in the following way. The spin-orbit mixing of ${}^4A_1$ states with the ground state in the first order approximation allows for its excitation when irradiated by an intense $2.8-3.0$~eV laser. The excited system then relaxes by first experiencing a non-radiative transition to ${}^2A_1$ followed by an optical emission at $\approx 2.05$~eV with an out-of-plane polarization.
We should add that due to the lack of orbital degeneracy, effective g-factor of electronic spin in ground and excited states are the same. Hence, in the presence of an external magnetic field spin-down and spin-up channels in the ground and excited states experience equal energy splitting. This means no Zeeman splitting in the photoluminescence spectrum of \vnnb\ should be expected~\cite{Li2017}.

Now, we turn to the discussion about the negatively charged \vb. We tentatively associate the electrical dynamics of this defect to the single-photon emitters observed in the labs, where there is a mismatch in the absorption and emission polarization of the defects (class-II). The laser largely excites the in-plane dipole moment and therefore the defect is excited to the $a_1'e'$ configuration. Then it either emits a linearly polarized photon, which could have a polarization in parallel or perpendicular to the absorption polarization, or with a finite probability that the color center emits photons with circular polarization. Therefore, the absorption and emission polarizations of the photons do not necessarily coincide. This indeed stems from the spatial degeneracy of the excited state. Our above theory is further supported by the \abinit results where the transition frequencies is expected to be about $1.77$~eV, which is in good agreement with the observed wavelengths of class-II emitters. This hypothesis is also corroborated by considering the processes induced by spin interactions. The non-radiative channels induced by the spin-orbit interaction shown in Fig.~\ref{fig:vbtrans}(b) are in agreement with the multi-channel transition reported in the experiments. The excited state ${}^3E'$ faces several competing relaxation processes. It could even rapidly get evacuated by a fast spin-orbit process owing to the small energy differences with the first excited state [see Table~\ref{tab:VB-energies} for the energy difference between the first, second, and third excited states].
It is noteworthy that the defect may even become excited to the $e'e''$ configuration via a photon with axial electric polarization and then decay to the first excited state through the non-radiative channels and finally relax by emitting an in-plane-polarized photon. In this case, the defect can exhibit a singlet-singlet transition as well. Such a transition is immune to the Zeeman splitting and magnetic field fluctuations as reported in a recent observation~\cite{Li2017}.
Interestingly, our study predicts an optical spin polarization channel in negatively charged boron vacancy in h-BN. A light beam at frequency around $1.77$~eV with linear polarization drives the system into ${}^3E'$, which will either relax back with a photon emission or will experience a non-radiative transition to the ground states with $m_S=\pm 1$. Since the linear polarization is more efficient in eviction of the $m_S=0$ state (see above discussion), one expects population transfer from $m_S=0$ to $m_S=\pm 1$ after a few optical circulations.

To summarize, the lowest calculated ZPLs are given in Table~\ref{tab:ZPLs} alongside the associated optical polarization of the emitted photons. In this table we also show the results for \vncb\ that are discussed in the supporting information.

\renewcommand{\arraystretch}{1.2}
\begin{table}[tb]
\caption{\label{tab:ZPLs} Optical zero phonon lines (ZPLs) and corresponding optical emission polarization (OEP) of charge neutral \vnnb, positively charged \vncb~[see supporting information], and negatively charged \vb\ defects as calculated by HSE $\Delta$SCF method in units of electronvolt. OEP notation is introduced after coordinate chosen in Fig.~\ref{fig:scheme}. The inplane clockwise and counter-clockwise circular polarizations are noted by $\hat x \pm i\hat y$.}
\begin{ruledtabular}
\begin{tabular}{lcc}
Defect & lowest ZPL & OEP \\
\hline
$[$\vnnb$]^0$ & 2.05 & $\hat{z}$ \\
\multirow{2}{*}{$[$\vb$]^-$} & 1.92\footnote{Room temperature; $D_{3h}$ excited state symmetry} & $\hat{x}\pm i\hat{y}$ \\
	& 1.77\footnote{Cryogenic temperatures; $C_{2v}$ excited state symmetry} & $\hat{x},\hat{y}$ \\
$[$\vncb$]^+$ & 1.51 & $\hat{z}$
\end{tabular}
\end{ruledtabular}
\end{table}
\renewcommand{\arraystretch}{1.0}
%

%
%
\section{Conclusion}
The group theoretical analysis as well as the calculated zero-phonon-lines have been exploited to relate the studied defects to the experimentally reported ZPLs, their excitation and emission profile of polarization, two-photon excitation processes, radiative and non-radiative channels of relaxation, and dark and bright shelving states. In particular, we have identified shelving states in the defects that can contribute in ISC processes. The ${}^4A_1$ state in \vnnb~is believed to play an important role in the optical dynamics of the defect and its existence can be experimentally verified by employing an exciting laser $\approx 3.0$~eV. The dark triplet ${}^3E_a''$ in the negatively charged \vb~has also been shown to induce competing non-radiative relaxations owing to its energetic proximity to the bright ${}^3E'$ state.
The spin-orbit interaction study has also allowed us to distinguish the observed multi-relaxation routes. Moreover, it predicts the low temperature electronic structure of the defects and identify an optical spin polarization channel for the \vnnb~and negatively charged \vb. 
Our work, therefore, is anticipated to shed more light on the road to identification of quantum emitters in h-BN monolayers. A better knowledge about those emitters, in turn, will considerably influence the nanophotonics and quantum technology.

%
%
\begin{acknowledgements}
This work was supported by the ERC Synergy grant BioQ, the EU EQUAM and DIADEMS projects, the DFG CRC TRR21, and DFG FOR 1493. Support from the Hungarian Na- tional Research Development and Innovation Office (NKFIH) in the frame of the Quantum Technology National Excellence Program (Project No. 2017-1.2.1-NKP-2017-00001) is acknowledged.
\end{acknowledgements}

%
%
\bibliography{origin}

\end{document}